\documentclass{article}
\usepackage{amssymb}
\usepackage{amsmath}

\newcommand{\code}[1]{\texttt{#1}}

\begin{document}

\title{Symbolic Script Programming for Java}

\author{
Raphael Jolly\\
{\small Databeans}\\
{\small Paris, France}\\
\texttt{raphael.jolly@free.fr}
\and
Heinz Kredel\\
{\small IT-Center, University of Mannheim}\\
{\small Mannheim, Germany}\\
\texttt{kredel@rz.uni-mannheim.de}
}

\maketitle

\begin{abstract}
Computer algebra in Java is a promising field of development. It has not yet
reached an industrial strength, in part because of a lack of good user
interfaces. Using a general purpose scripting language can bring a natural
mathematical notation, akin to the one of specialized interfaces included in
most computer algebra systems. We present such an interface for Java
computer algebra libraries, using scripts available in the JSR 223 framework.
We introduce the concept of `symbolic programming' and show its usefulness
by prototypes of symbolic polynomials and polynomial rings.
\end{abstract}

%
%

\section{Introduction}

Computer algebra in Java has many advantages over implementations in
more rustic languages such as Lisp or C, and was shown to be
reasonably fast compared to them
\cite{Kredel:2006,Kredel:2007,Kredel:2008}. However it lacks good user
interfaces.  The idea of using a general purpose scripting language as
glue code and interface to C-libraries and standalone computer algebra
systems was introduced by the Sage project \cite{Stein:2005} with its
Python interface \cite{vanRossum:1991}. The fact that this solution
brings a natural mathematical notation, was key in this project's
impressive development. We discuss how such interfaces can be build in
the case of Java computer algebra libraries. We are considering
scripts available in the JSR 223 framework \cite{JSR223:2006}, and
also Scala \cite{Odersky:2003} whose interpreted mode is not formally
part of JSR 223 although this is planned.

\subsection{Background and related work}


We have started this discussion with the paper \cite{JollyKredel:2008}
where we have shown that at least the scripting languages Ruby, Groovy,
Python and Scala are suitable as domain specific languages for computer
algebra \cite{Matsumo:1995,JRuby:2003,Jython:1997,Groovy:2003}.  In this
paper we have described the following required features: object
orientation, operator overloading (in particular the availability of good
operators for writing rational numbers and powers), and coercions (either
by double-dispatch, base types redefinition, or implicit conversion).
For each of the assessed scripting languages we have sketched a way to
enter algebraic expressions and to transfer them to desired Java objects.
Since then we have implemented most of these ideas in a Jython front-end
to our two computer algebra libraries, JAS \cite{Kredel:2000} and ScAS.

The former is a new approach to computer algebra software design and
implementation in an object oriented programming language.  It
provides a well designed software library using generic types for
algebraic computations implemented in the Java programming language.
The mathematical focus of JAS is at the moment on commutative and
solvable polynomials, Gr\"obner bases, an experimental factorization
package and applications. JAS has a Jython interface and also a
prototypical Groovy interface.

The latter (ScAS) is a reimplementation in Scala of JSCL \cite{Jolly:2007}, which
was a pre-Java 5, non Generics attempt at a ``Java symbolic computation library''.
The new, generic type based implementation is inspired by the ideas developed in
JAS, and is also meant as an exploration of the specific features of Scala. We
are making access to these libraries with a selection of scripts. In its current
state, ScAS can be accessed both by Scala in interpreted mode (which makes a
nice, uniform solution), and Jython. JSCL was accessed with BeanShell
\cite{BeanShell:2000}.

There are other projects to implement a computer algebra system completely in
Python: SymPy \cite{Certik:2008} or to provide user interfaces based on the
Eclipse Rich Client Platform \cite{Eclipse:2008} as MathEclipse
\cite{Kramer:2008} does. For other related work see the discussion and
references in \cite{JollyKredel:2008}.

\subsection{Outline of the paper}

In this paper we extend the concepts of \textit{input} of algebraic expressions
in a scripting language to the \textit{output} of these expressions such that
they can be reused as input in further computations.  This is done in section 2
with an introduction into the concept of \textit{Symbolic programming}. The
main part of the paper in section 3 presents prototypes of symbolic polynomials
and polynomial rings. For polynomial rings we partially solve the problem of
defining objects representing algebraic structures. Finally, section 4
concludes.

\section{Symbolic Programming}

In this section we describe and formalize a new meta-programming technique
baptized ``symbolic programming''. This technique is applicable in any object
oriented language, but is especially interesting in the case of interpreted
languages, where the output of an instruction can be reused and combined in
further statements to the interpreter, interactively. To our knowledge, it was
first used on a significant scale, albeit not under this name, in Sage.

\subsection{Reconstructing expression}

We define the following items:

Definition 1.
In a given language, an expression is said to be \textit{reconstructing} iff:
(i) it is a valid expression of said language, and
(ii) it has same input and output forms, in the sense of string equality.

The output form is as obtained for example by the \code{to\-String()} method in
Java. Expression reconstruction will usually depend on the actual context of the
computation. So a context needs to be taken into account, since there are cases
where, for the expression to be valid, itself or its constituents have to be
defined in the sense of the programming language.  

Definition 2.
A \textit{context} $c$ is a set of definitions of variables or functions of the
form $n(X_i)$, with $(X_i)$ a (possibly empty) list of parameters, such that in
$c$, if $x_i$ is reconstructing for all $i$, then $n(x_i)$ is reconstructing,
i.e. $n(x_i)$ has output form $n(x_i)$.

The next definition is central to the concept of reconstructing expressions.

Definition 3.
In a given language, an expression is said to be
\textit{reconstructing in context $c$} iff:
(i) it is a valid expression of said language, taking $c$ into account, and
(ii) it has same input and output forms.

\subsection{Symbolic object, symbolic type}

We further define:

Definition 4.
In a given language and context $c$, an object is called a
\textit{symbolic object} iff its output form is reconstructing in
$c$, as defined in 3.

Definition 5.
In a given language, a type $t$ is called a \textit{symbolic type} iff for all
instances $o$ of $t$ there exists a context where $o$ is symbolic.

Here are some first examples.


Java's type \code{int} is symbolic because the output form of,
say \code{1} is \code{1}, which is the same as the original expression.

Note. Symbolic programming as understood in this trivial sense is available in
any known programming language, not just object oriented languages. More
interesting examples follow.


The Java class \code{String} is \textit{not} symbolic because
\code{"hello world".toString()} yields \code{hello world},
which is not the same as \code{"hello world"}.

Contrarily, Scala's \code{Symbol} class is symbolic, because \code{'a} yields
\code{'a}.


Java's type \code{long} is not symbolic because of the `l' or `L' that
must be appended to literals and gets removed when the value is printed.

\subsection{Container types: array, collections, tuples}

A Java array, as defined for example in

\begin{verbatim}
  int n[] = new int[] {1, 2};
\end{verbatim}

is not symbolic, because it yields a unique identifier, not its contents.
Scala's class \code{List[A]}, however, is symbolic (if \code{A} is), as shown
below.

Definition 6.
A container type $q$ is called a \textit{symbolic container type} iff for any
instance $p$ of $q$, if all the components of $p$ are symbolic, then $p$
is symbolic.

One can show that a symbolic container type $q$ is a symbolic type if its
component type(s) is (are) symbolic.

Example : Scala's class \code{List[A]} is a symbolic container type, since
its instances are symbolic when their components are: \code{List(1, 2)} yields
\code{List(1, 2)}. Hence, if A is a symbolic type, then so is \code{List[A]}.

\subsection{Applications}

\subsubsection*{Big integers}
\label{sec:big}

The class \code{BigInteger} of Java is not a symbolic class, as its
output form becomes an invalid Java expression when the value
exceeds integer capacity. Hence we have to define our own symbolic BigInt
class. As an output form, we can choose: \code{new BigInt("1")}.
The resulting code is:

\begin{verbatim}
import java.math.BigInteger;

public class BigInt {
  BigInteger value;

  public BigInt(String val) {
    value = new BigInteger(val);
  }

  public String toString() {
    return "new BigInt(\""+value.toString()+"\")";
  }
}
\end{verbatim}

The \code{toString()} produces the desired \code{new BigInt("... ")},
which in turn is a valid constructor expression to get the big integer
back.

\subsubsection*{Fractions}
\label{sec:fra}

We want to define a Rational class that enables us to encode e.g.
$ \frac{1}{2}$. Regarding the output form, we would
ideally have indeed \code{1/2}, but no scripting language exists that won't
interpret this as an int or float. In some of them (Ruby, Groovy), the divide
operator can be overloaded not to execute the division, but then we can't use it
for integer division anymore. Some other scripting languages allow to define and
use an operator such as \code{1//2}.

In Python, an appealing syntax is with a Tuple \code{(a, b)} or List
\code{[a, b]}. Indeed these types are symbolic container types, so that
\code{(1, 2)} yields \code{(1, 2)}. It should be noted however that this
notation is limited in its operations, for one could not write
\code{(1, 2)+(2, 3)} and expect the rational sum as a result.

To improve on this, we consider notations such as \code{frac(a, b)}. In Java,
we can populate the context with the following definition:

\begin{verbatim}
  Rational frac(int n, int d) {
    return new Rational(n, d);
  },
\end{verbatim}

and define the following class:

\begin{verbatim}
public class Rational {
  int numerator;
  int denominator;

  public Rational(int n, int d) {
    numerator = n;
    denominator = d;
  }

  public String toString() {
    return "frac("+n+", "+d+")";
  }
}
\end{verbatim}

Alternatively we could avoid using contexts and define an output
form as \code{new BigRat(1,2)} or \code{new BigRat ("1","2")},
like the \code{BigInt} above.

\subsubsection*{Powers}

There are several possible operators in the various scripting languages. In
Python, Ruby, Groovy, \code{a**2} is available. In Scala, it is not possible
because \code{**} has the same precedence as \code{*}; there are other
characters with a higher precedence like \verb/\/, but there is another problem:
unary operators such as \code{-} have a even higher precedence, which means that
\verb/-x\2/ becomes \verb/x\2/. In all the former languages and in Java, one
can use a notation similar to that used above for fractions: \code{pow(a, 2)}.

\subsection{Type preservation}

In \ref{sec:big} we have defined a syntax that is a bit clumsy for big integers,
perhaps we could note these like regular integers for small enough values. For
this to happen, their type will have to mutate.

Definition 7.
A symbolic type is said to \textit{mutate} iff some of its instances have an output
form that is an expression with a different type than their input form.

Definition 8.
The types reachable from a class A through transitive closure of the relation
``type A mutates to type B'' form the \textit{symbolic type group} of A.

Definition 9.
A symbolic type whose symbolic type group contains a single element (itself)
is said to be \textit{preserved}.

Our \code{BigInt} example above becomes:

\begin{verbatim}
public class BigInt {
...
  public String toString() {
    return value.bitLength()<32 ?
      value.toString() :
      "new BigInt(\""+value.toString()+"\")";
  }
}
\end{verbatim}

The resulting symbolic type group is the set (int, BigInt).

\subsection{Type preservation examples}

\subsubsection*{Big Integers}

In Python, the set (PyInteger, PyLong) is a symbolic type group. One has to note
that these types have a uniform syntax: no signalization is added when 32 bit
precision is exceeded (and there is no further limit since PyLong is the
equivalent of Java's BigInteger), which is much nicer than what we proposed
above for the Java case.

\subsubsection*{Polynomials}

This is in slight anticipation to section \ref{sec:pol}, where the
polynomial case is discussed in detail. Suppose we define a Polynomial
type, with the following syntax: \code{1+x}. When, for example, we
subtract \code{x}, the type mutates, and we get \code{1}, which is of
type int. If we want the Polynomial type to be preserved, we can
insert a conversion like \code{x.valueOf(1)}. Alternatively, we can
use an unsimplified syntax like \verb/x^0/, which is still of type
Polynomial. In both cases, we don't get a natural mathematical
notation as we aim, so we will have to use a mutating type. Note:
type preservation is context-dependent, as shown in the next example.

\subsubsection*{Rationals}

If we define a Rational type with the syntax \code{frac(1, 2)}, then
depending on the return type of the \code{frac} method, the type could
either stay the same as in our application in \ref{sec:fra}, or it could
become a floating point if the result is defined to be $0.5$ (as we
would do if indeed we want a numeric evaluation).

\subsubsection*{ModIntegers}

ModIntegers can carry the modulo information with them:
$ 5 \mod 11$ can be noted \code{mod(5, 11)}. But we can also
remove the modulo without causing any error. On the other hand, symmetric
(negative) modular numbers, do need the information because otherwise the value
will be wrong: $ -5 \equiv 6 \mod 11$ which is not the same
as $ -5$. In fact, it depends what other objects it is
supposed to interact with: if all numbers are given
$ \mod 11$, then we get correct results. In general, a type
must be enabled for interaction with all the members of its symbolic type group,
and reciprocally.

\section{A Symbolic Polynomial type}
\label{sec:pol}

Polynomials are ubiquitous in computer algebra. They allow to manipulate
unknowns such as $x$ as if they were numbers, which is the essence of symbolic
computation. A polynomial is defined over a base ring (a set of elements which
can be added, subtracted, multiplied and which has a zero and a one) and is
itself a ring element. Therefore, we must enable both our base type and our
polynomial type for these ring operations.

\subsection{Library class definitions}
\label{sec:libclass}

In our backing Java library we assume for example the following classes.
\code{BigInt} is our prototype for base coefficients and \code{Polynomial}
is the main polynomial class. For a complete list of currently available
base coefficients and polynomial implementations see our online documentation
at \cite{Kredel:2000,Jolly:2007}. The backing Java class library can be
implemented directly in Java as in JAS or indirectly via Scala, which
generates Java byte code, as in ScAS.

\begin{verbatim}
class BigInt {
  BigInt add(BigInt that);
  BigInt subtract(BigInt that);
  BigInt multiply(BigInt that);
}
\end{verbatim}

\begin{verbatim}
class Polynomial {
  Polynomial add(Polynomial that);
  Polynomial subtract(Polynomial that);
  Polynomial multiply(Polynomial that);
  Polynomial pow(BigInt exp);
}
\end{verbatim}

We further assume that polynomials are created using polynomial
factories in the library.

\begin{verbatim}
Polynomial x =
  new PolynomialFactory(
    new BigInt(), new String[] {"x"})
      .generator(0)
\end{verbatim}

This design is meant to separate the informations about the nature of
the ring from those which regard the operations on its elements
\cite{Niculescu:2003, Niculescu:2004}.  It further turns abstract
mathematical entities like polynomial rings to first class citizens of
the programming language.

\subsection{Abstract coefficient type}

In JAS the classes just defined inherit from an abstract type, using F-bounded
polymorphism \cite{Cook:1989} as available in Java since version 5 (JDK 1.5):

\begin{verbatim}
interface Ring<T extends Ring<T>> {
  T add(T that);
  T subtract(T that);
  T multiply(T that);
},
\end{verbatim}

which allows to define a generic polynomial type:

\begin{verbatim}
class Polynomial<C extends Ring<C>> implements
    Ring<Polynomial<C>> {
  Polynomial<C> add(Polynomial<C> that);
  Polynomial<C> subtract(Polynomial<C> that);
  Polynomial<C> multiply(Polynomial<C> that);
  Polynomial<C> pow(BigInt exp);
}
\end{verbatim}

This polynomial class can have any ring class as its base ring, including
itself, which is quite powerful, but poses a challenge regarding our aimed
natural mathematical notation. This is where scripting comes into play. Below
we exemplify with Jython, but other scripts are possible.

\subsection{Coercion of coefficients}

A polynomial must be able to be added, subtracted etc. (to) its
coefficients, which are part of its symbolic type group (i.e. the
polynomial type can mutate to the type of its coefficients), and
reciprocally. We could define ad-hoc methods such as:

\begin{verbatim}
  Polynomial<C> add(C coef);
  Polynomial<C> subtract(C coef);
  Polynomial<C> multiply(C coef);
\end{verbatim}

The first problem is with inversed operands, when the coefficient
comes first.  As we've seen in \cite{JollyKredel:2008}, section 3.1,
in Python this is addressed by double-dispatch. A second problem is
that when the base ring is itself a polynomial ring, we must accept
not only coefficients but coefficients of coefficients, and so on,
which rules this technique out. Instead, we can coerce coefficients
to the polynomial type, using a method such as:

\begin{verbatim}
  Polynomial<C> valueOf(C coef);
\end{verbatim}

Then, to add $1$ to $ y \in \mathbb{Z}[x][y]$, we will have to
write: \code{y.add(y.valueOf(x. valueOf(1)))}. Again, scripting will come to help
and make this coercions for us silently. This is possible because, unlike Java,
Python is dynamically typed and can use the same method to add $y$ or $1$. In
said method, some \code{isinstance()} calls check the type of the argument and
perform the required nested calls to \code{valueOf()}.

We could emulate dynamic typing in Java, using method arguments of type
\code{Object} (or any other relevant superclass) and then check the run-time
type with \code{instanceof} and make type casts. In fact, this is what we did in
the former version of JAS and JSCL. But Java 5 Generics polymorphism allows a
much safer design, where adding apples to oranges is forbidden at compile time.
The downside is that the entailed conversions are to the user's burden.
Scripting comes in to solve this problem and enables the best of both worlds:
type safety for the lower layers with an easy to use scripting interface.

\subsection{Output form}

We would like our polynomial to be a symbolic container type for its
coefficients (see definition 6), such that
\code{x.pow(2).add(x.multiply(2)).add(1)} is reconstructing. But this is
not quite yet a natural mathematical notation, and we need operator overloading
\cite{JollyKredel:2008}. It enables such output form as
\code{x**2+2*x+1}, valid in several scripting languages, including Python (but
not BeanShell for instance).

\subsection{Implementation}

A prototype of these ideas is given below. Note that the coercion of the
coefficients is performed not by the polynomial itself, but by a factory. This
design reflects the design of the Java class libraries as explained in section
\ref{sec:libclass}. Our Jython scripting interface using the ScAS API is
implemented as follows.

Class \code{Ring} is the main scripting interface for polynomial factories.

\begin{verbatim}
class Ring:
  def __init__(self,ring,vars,\
      ordering=Lexicographic):
    self.ring = PolynomialFactory(ring,\
        vars, ordering)

  def __str__(self):
    return str(self.ring)

  def gens(self):
    return [RingElem(x) for x in\
        self.ring.generators()]
\end{verbatim}


The method named \verb/__init__/ is the constructor in Python and method
\verb/__str__/ has fixed meaning in Python, which is similar to Java's
\code{toString()} method.  Our method \code{gens()} returns a list of
generators for the respective polynomial ring. The generators are actual
Java objects in the respective library wrapped by the scripting
\code{RingElem} class.

\begin{verbatim}
def lift(factory,p):
  if not factory.equals(p.factory()):
    p = factory.valueOf(lift(factory.ring(),p))
  return p
\end{verbatim}

The standalone \code{lift()} method above is used to lift a coefficient to an
element of the given polynomial ring. The following \code{RingElem} is the main
scripting interface for polynomials.

\begin{verbatim}
class RingElem:
  def __init__(self,elem):
    self.elem = elem

  def __str__(self):
    return self.elem.toString()

  def __abs__(self):
    return RingElem(self.elem.abs())

  def __neg__(self):
    return RingElem(self.elem.negate())

  def __mul__(self,other):
    (s,o) = self.coerce(other)
    return RingElem(s.elem.multiply(o.elem))

  def __rmul__(self,other):
    (s,o) = self.coerce(other)
    return RingElem(o.elem.multiply(s.elem))

# same for add, sub, etc

  def __pow__(self,exp):
    return RingElem(self.elem.pow(int2bigInt(exp)))

  def __eq__(self,other):
    (s,o) = self.coerce(other)
    return s.elem.equals(o.elem)

  def __ne__(self,other):
    (s,o) = self.coerce(other)
    return not s.elem.equals(o.elem)
\end{verbatim}

As mentioned already, the method with leading and trailing underscores
in their names have predefined meaning in Python, which is easy to
guess: \verb/__mul__(.)/ implements the multiplication operator
`\code{*}' of Python, etc. Our implementation of these methods just
delegate to the respective Java object method invocations.

\begin{verbatim}
  def coerce(self,other):
    base = self.base()
    if isinstance(base,BigInt):
      if isinstance(other,PyInteger):
        other = RingElem(int2bigInt(other))
      elif isinstance(other,PyLong):
        other = RingElem(long2bigInt(other))
    if self.depth() < other.depth():
      return (other.lift(self),other)
    else:
      return (self,self.lift(other))

  def lift(self,other):
    return RingElem(lift(self.factory(),\
        other.elem))

  def depth(self):
    n = 0
    r = self.factory()
    while isinstance(r, PolynomialFactory):
      n += 1
      r = r.ring()
    return n

  def base(self):
    r = self.factory()
    while isinstance(r, PolynomialFactory):
      r = r.ring()
    return r

  def factory(self):
    return self.elem.factory()
\end{verbatim}


Our method \code{coerce()} has to adjust the types of this element and
the other element. It uses the methods \code{base()}, \code{depth()},
\code{lift()} and \code{factory()} to convert Python objects to
\code{RingElem} objects and then to adjust and coerce types, so that
the \verb/__X__/ operations are well defined.

\subsection{Sample session}

Using our interface is as simple as adding the library to the classpath, running
Jython, and making some imports. As desired the polynomials are symbolic, since
for example the polynomial expression \code{x+x*y} is printed exactly as
\code{x+x*y}.

\begin{verbatim}
 Jython 2.2.1 on java1.6.0_11
 Type "copyright", "credits" or "license" for more\
 information.
 >>> from interface import Ring, BigInt
 >>> r = Ring(BigInt(), ["x"])
 >>> print(r)
 ZZ[x]
 >>> [x] = r.gens();
 >>> print(1+x)
 1+x
 >>> [y] = Ring(x.factory(), ["y"]).gens()
 >>> print(x+x*y)
 x+x*y
 >>> [z] = Ring(y.factory(), ["z"]).gens()
 >>> print((1-z)**2)
 1-2*z+z**2
 >>> print(z.factory())
 ZZ[x][y][z]
\end{verbatim}

Note, the ring objects are not symbolic, as \code{Ring(BigInt (), ["x"])}
has output form \code{ZZ[x]}, which is not reconstructing.

\subsection{Symbolic ring factory}
\label{sec:ringFac}

Our algebraic structures, for example polynomial rings, are
represented by factory classes and instantiated objects in the Java
and Scala libraries.  This design is also followed in the scripting
interface. For these objects definition 4 applies and we
look for a scripting interface to these objects which makes them
symbolic objects.  In the following we sketch an implementation of a
symbolic polynomial ring factory.  The sample code is based on the JAS
API.

\begin{verbatim}
class PolyRing(Ring):
  def __init__(self,coeff,vars,order):
    if coeff == None:
        raise ValueError, "No coefficient."
    cf = coeff;
    if isinstance(coeff,RingElem):
        cf = coeff.elem.factory();
    if isinstance(coeff,Ring):
        cf = coeff.ring;
    if vars == None:
        raise ValueError, "No variabls given."
    names = vars;
    if isinstance(vars,PyString):
        names = StringUtil.variableList(vars);
    nv = len(names);
    to = PolyRing.lex;
    if isinstance(order,TermOrder):
        to = order;
    ring = GenPolynomialRing(cf,nv,to,names);
    self.ring = ring;
\end{verbatim}

\begin{verbatim}
  def __str__(self):
    cf = self.ring.coFac;
    cfac = cf;
    if cf.equals( BigInteger() ):
        cfac = "ZZ()";
    # ...
    if cf.getClass() == \
        GenPolynomialRing(BigInteger(),1)
        .getClass():
        cfac = str(PolyRing(cf.coFac,\
                   cf.varsToString(),cf.tord));
    to = self.ring.tord;
    tord = to;
    if to.evord == TermOrder.INVLEX:
        tord = "PolyRing.lex";
    if to.evord == TermOrder.IGRLEX:
        tord = "PolyRing.grad";
    nvars = self.ring.varsToString();
    return "PolyRing(%s,%s,%s)" %
            (cfac, "\""+nvars+"\"", tord);
\end{verbatim}

\begin{verbatim}
  lex = TermOrder(TermOrder.INVLEX)
  grad = TermOrder(TermOrder.IGRLEX)
\end{verbatim}

The \verb/__init__/ constructor takes the main constituents of a
polynomial ring as parameters: \code{coeff}, a factory for
coefficients, \code{vars}, the names of the variables and
\code{order}, the desired term order.  From this information we
assemble the required parameters for the library polynomial ring
constructor \code{Gen\-Poly\-nomial\-Ring}.  There are two constants
\code{lex} and \code{grad} which represent term orders as defined in
the library. The method \verb/__str__/ assembles the string parts from
the library ring object and returns a string, which is identical to
the \code{PolyRing} expression.  The low level tests
\code{cf.equals(...)} or \code{cf.getClass()} and the explicit
construction of the output form will be replaced in the future
by a method \code{cf.toScript()}. This method will be similar to the
usual \code{cf.toString()} except it will yield an output form which
is reconstructing in the respective scripting language.

A sample session using this \code{PolyRing} code looks as follows.


\begin{verbatim}
 >>> r = PolyRing(ZZ(),"B,S",PolyRing.lex);
 >>> print "Ring: " + str(r);
 Ring:   PolyRing(ZZ(),"B,S",PolyRing.lex)
\end{verbatim}

Which shows that the defined ring is symbolic. This construction works
also for recursive polynomial rings using \code{r} as coefficient factory.

\begin{verbatim}
 >>> pr = PolyRing(r,"T,Z",PolyRing.lex);
 >>> print "PolyRing: " + str(pr);
PolyRing: PolyRing(
              PolyRing(ZZ(),"B,S",PolyRing.lex),
                      "T,Z",PolyRing.lex)
\end{verbatim}

\code{ZZ()} denotes the class corresponding to \code{BigInt} above
which must be available in the context.

\section{Conclusion}

We have extended the work outlined in our previous article
\cite{JollyKredel:2008}, where we studied the suitability of four scripting
languages for symbolic algebraic computation. In this paper we focused our
work on only one scripting language, namely Python with its Java implementation
Jython. However, we expect that our work can also be transfered at least to Ruby
(JRuby). Groovy was not further considered because we had to use either ``use''
blocks, with limited interactivity, or \code{ExpandoMetaClass}es which require
to redefine base types in low-level Java. Our view of Scala has shifted a bit
away from a scripting language to a language suited for library development as
well.

We have shown how to design a scripting interface for Java computer algebra
implementations which satisfies the concept of \textit{reconstructing expressions}.
This concept makes it possible to reason precisely about various
parts of the user interface for computer algebra systems backed by elaborated
programming libraries. We can handle the input of polynomial expressions and
have precise requirements for the output of the algebraic transformations.
This goes far beyond the capabilities of our Java systems before, as they were
limited to custom parsers and output routines. We now have also some concept
for the representation of ring factories in the scripting language as symbolic
objects. To demonstrate our ideas we designed a symbolic polynomial with a
prototype implementation in Jython as front-end to our Java libraries JAS
and ScAS.

\subsection*{Acknowledgments}

We thank our colleagues for various discussions and for encouraging our work.

\bibliographystyle{abbrv}


\end{document}